\documentclass[aps,prl,twocolumn,groupedaddress]{revtex4-1}
\usepackage{graphicx}
\usepackage{bm}
\usepackage{amsmath}
\usepackage{amssymb,wasysym}

\begin{document}


\title{Revealing the Superfluid Lambda Transition in the Universal Thermodynamics of a Unitary Fermi Gas}


\author{Mark J. H. Ku}
\author{Ariel T. Sommer}
\author{Lawrence W. Cheuk}
\author{Martin W. Zwierlein}

\affiliation{Department of Physics, MIT\textendash Harvard Center for Ultracold Atoms, and Research Laboratory of Electronics, Massachusetts Institute of Technology, Cambridge, MA 02139, USA}


\date{\today}

\begin{abstract}
We have observed the superfluid phase transition in a strongly interacting Fermi gas via high-precision measurements of the local compressibility, density and pressure down to near-zero entropy. Our data completely determine the universal thermodynamics of strongly interacting fermions without any fit or external thermometer. The onset of superfluidity is observed in the compressibility, the chemical potential, the entropy, and the heat capacity. In particular, the heat capacity displays a characteristic lambda-like feature at the critical temperature of $T_c/T_F = 0.167(13)$. This is the first clear thermodynamic signature of the superfluid transition in a spin-balanced atomic Fermi gas.
Our measurements provide a benchmark for many-body theories on strongly interacting fermions, relevant for problems ranging from high-temperature superconductivity to the equation of state of neutron stars.

\end{abstract}


\maketitle


Phase transitions are ubiquitous in Nature: water freezes into ice, electron spins suddenly align as materials turn into magnets, and metals become superconducting. A pervasive feature of continuous phase transitions is their critical behavior, namely singularities in thermodynamic quantities: the magnetic susceptibility diverges at a ferromagnetic transition, the specific heat shows a jump at superconducting transitions~\cite{tink04sc} as well as at the superfluid transition of $^3$He~\cite{voll90}. In $^4$He, at the famous $\lambda$-transition into the the superfluid state, the jump is even resolved, in zero gravity, to be a near-diverging, singular peak~\cite{lipa03lambda}. A novel form of superfluidity has been realized in trapped, ultracold atomic gases of strongly interacting fermions, particles with half-integer spin~\cite{kett08rivista,gior08review,bloc08review}. Thanks to an exquisite control over relevant system parameters, these gases have recently emerged as a versatile system well suited to solve open problems in many-body physics~\cite{bloc08review}. However, while superfluidity has been established via the observation of vortex lattices in rotating gases~\cite{zwie05vort}, no clear thermodynamic signature of the superfluid transition has previously been observed.

Initial measurements on the thermodynamics of strongly interacting Fermi gases have focused on trap averaged quantities~\cite{kina05heat,stew06pot,luo09thermo} in which the superfluid transition is inherently difficult to observe. It is also challenging to reveal the critical behavior through the study of local thermodynamic quantities. The emergence of the condensate of fermion pairs in a spin-balanced Fermi gas is accompanied by only minute changes in the density~\cite{kett08rivista}. Therefore, quantities that involve integration of the density over the local potential, such as the energy $E$~\cite{hori10thermo} and the pressure $P$~\cite{nasc10thermo}, are only weakly sensitive to the sudden variations in the thermodynamics of the gas that one expects near the superfluid phase transition~\cite{ku11supporting}.

A thermodynamic quantity involving the second derivative of the pressure $P$ is expected to become singular at the second order phase transition into the superfluid state. An example is the (isothermal) compressibility $\kappa = \frac{1}{n}\frac{\partial n}{\partial P}|_{T}$, the relative change of the gas density $n$ due to a change in the pressure $P$. As the change in pressure is related to the change in chemical potential $\mu$ of the gas via ${\rm d}P = n \,{\rm d}\mu$ at constant temperature, $\kappa = \frac{1}{n^2}\frac{\partial^2 P}{\partial\mu^2}|_T$ is a second derivative of the pressure, and thus should reveal a clear signature of the superfluid transition.

In the present work, we report on a high-precision measurement of the local compressibility, density and pressure across the superfluid phase transition in the unitary Fermi gas, realized using a trapped, two-spin mixture of fermionic $^6$Li atoms at a Feshbach resonance~\cite{kett08rivista,bloc08review}. The combination of these three directly measurable, local quantities determines the entire thermodynamics of the homogeneous gas as we show below. Thermometry of strongly interacting Fermi gases was known to be notoriously difficult. Our method solves this problem. It is general and applies to other systems as well, such as Bose gases in three or two dimensions.

In order to determine the thermodynamic properties of a given substance, the general goal is to measure an equation of state (EoS), such as the pressure $P(\mu,T)$ as a function of the chemical potential $\mu$ and the temperature $T$. Equivalently, replacing the pressure by the density $n = \frac{\partial P}{\partial \mu}|_T$, one can determine the density EoS $n(\mu,T)$.

In our experiment we directly measure the local gas density $n(V)$ as a function of the local potential $V$ from in-situ absorption images of a trapped $^6$Li gas. The trapping potential is cylindrically symmetric, with harmonic confinement along the axial direction. This symmetry allows to find the 3D density via the inverse Abel transform of the measured column density~\cite{shin06phase}. Other than cylindrical symmetry, no other assumption on the shape of the potential is made. Instead, the local potential is directly measured via the atomic density distribution and the accurately known harmonic potential along the axial direction of the atom trap.

The compressibility $\kappa$ follows as the change of the density $n$ with respect to the local potential $V$ experienced by the trapped gas. The change in the local chemical potential is given by the negative change in the local potential, ${\rm d}\mu = -{\rm d}V$, and hence the compressibility is $\kappa = -\frac{1}{n^2} \frac{\mathrm{d} n}{\mathrm{d} V}|_T$.

The compressibility allows replacing the unknown chemical potential $\mu$ in the density EoS $n(\mu,T)$ by its known variation in the atom trap, yielding the equation of state $n(\kappa,T)$.
In lieu of the a priori unknown temperature, we determine the pressure $P(V) = \int_{-\infty}^\mu {\rm d}\mu' n(\mu') = \int_V^\infty {\rm d}V' n(V')$ that is simply the integral of the density profile over the potential~\cite{chen07,ho10bulk}. The resulting equation of state $n(\kappa,P)$ contains only quantities that can be directly obtained from the density distribution. This represents a crucial advance over previous methods that require the input of additional thermodynamic quantities, such as the temperature $T$ and the chemical potential $\mu$, whose determination requires the use of a fitting procedure or an external thermometer, as in~\cite{hori10thermo, nasc10thermo}.

We normalize the compressibility and the pressure by the respective quantities at the same local density for a non-interacting Fermi gas at $T=0$, $\kappa_0 =\frac{3}{2} \frac{1}{n E_F}$ and $P_0 = \frac{2}{5} n E_F$, where $E_F = \frac{\hbar^2 (3\pi^2 n)^{2/3}}{2m}$ is the Fermi energy and $m$ is the particle mass, yielding $\tilde{\kappa}\equiv\kappa/\kappa_0$ and $\tilde{p}\equiv P/P_0$.
In general, $\tilde{\kappa}$ could depend on dimensionless quantities other than $P/P_0$, such as the interaction parameter $n a^3$ of the gas, where $a$ is the scattering length. However, for dilute gases at the Feshbach resonance, the scattering length diverges and is no longer a relevant length scale. In the absence of an interaction-dependent length scale, the thermodynamics of such resonant gases are universal~\cite{ho04uni}, and $\tilde{\kappa}$ is a universal function of $\tilde{p}$ only.
Every experimental profile $n(V)$, whatever the trapping potential, the total number of atoms or the temperature, must produce the same universal curve $\tilde{\kappa}$ versus $\tilde{p}$. By averaging many profiles, one obtains a low-noise determination of $\tilde{\kappa}(\tilde{p})$.

Our method has been tested on the non-interacting Fermi gas that can be studied in two independent ways: In spin-balanced gases near the zero-crossing of the scattering length and in the wings of highly imbalanced clouds at unitarity, where only one spin state is present locally. Both determinations yield the same non-interacting compressibility EoS.

\begin{figure}[!h]
    \centering
    \includegraphics[width=90mm]{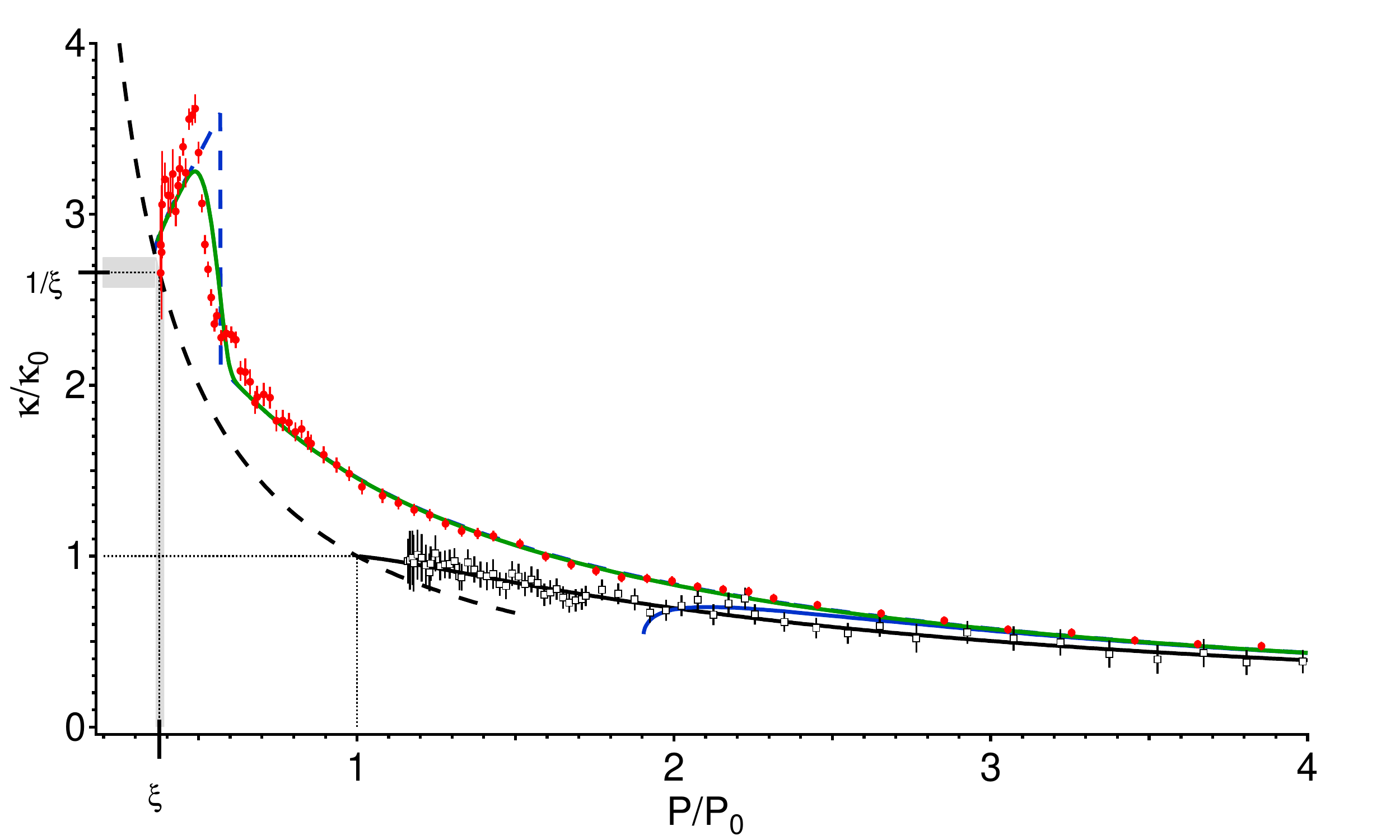}
    \caption{\label{fig1} Compressibility $\kappa$ versus pressure $P$ of the unitary Fermi gas. Red solid dots show $\kappa$ and $P$ normalized by their respective values $\kappa_0$ and $P_0$ for a non-interacting Fermi gas at the same density. The error bars show one standard deviation. The blue solid line shows the 3rd order Virial expansion. Black open squares (black solid line) denote data (theory) for a non-interacting Fermi gas. The black dashed line shows the relation $\tilde{\kappa} = 1/\tilde{p}$ that must be obeyed at zero-temperature both for the non-interacting gas ($\tilde{\kappa} = 1/\tilde{p} = 1$) and the unitary gas ($\tilde{\kappa} = 1/\tilde{p} = 1/\xi$) (dotted lines). A gray band marks the uncertainty region for the $T=0$ value of $\tilde{\kappa}= 1/\xi$ and $\tilde{p} = \xi$. The blue dashed curve is a model for the EoS of the unitary Fermi gas (above $T_c$: interpolation from the Monte-Carlo calculation~\cite{vanhouke11crossval}, below $T_c$: BCS theory including phonon and pair-breaking excitations). Green solid line: Effect of $2\mu{\rm m}$ optical resolution on the model EoS.}
\end{figure}

Fig.~1 shows the resulting compressibility equation of state $\tilde{\kappa}(\tilde{p})$. In the high-temperature and, equivalently, low-pressure regime, the pressure (and hence all other thermodynamic quantities) allows for a Virial expansion in terms of the fugacity $e^{\beta\mu}$~\cite{liu09virial}: $P \beta\lambda^3 = 2\sum_j b_j e^{j\beta\mu}$, with the $n$th order Virial coefficients $b_n$. It is known that $b_1=1$, $b_2=3\sqrt{2}/8$~\cite{beth37virial}, and $b_3=-0.29095295$~\cite{liu09virial}. Our data agrees excellently with the Virial expansion. Fixing $b_2$ and $b_3$, our measurement yields a prediction for $b_4 = +0.065(10)$, in agreement with ~\cite{nasc10thermo}, but contradicting a recent four-body calculation that gives a negative sign~\cite{raks11virial}.

At degenerate temperatures, the normalized compressibility rises beyond that of a non-interacting Fermi gas, as expected for an attractively interacting gas. A sudden rise of the compressibility at around $\tilde{p} = 0.55$, followed by a {\it decrease} at lower temperatures marks the superfluid transition. Indeed, the theory of superfluid phase transitions in three dimensions implies a singularity of the compressibility at the transition. This would appear as a sudden change in slope of $n(V)$ that is rounded off by the finite resolution of our imaging system.
Below the transition temperature, the decrease of the compressibility is consistent with the expectation from Bardeen-Cooper-Schrieffer (BCS) theory where single-particle excitations freeze out and pairs form (see model in Fig.~1).

As $T\rightarrow 0$, the Fermi energy $E_F$ is the only intensive energy scale, and so the chemical potential must be related to $E_F$ by a universal number, $\mu = \xi E_F$, where $\xi$ is known as the Bertsch parameter~\cite{bloc08review,gior08review}. It follows that at $T=0$, $\tilde{\kappa} = 1/\tilde{p}=1/\xi$. The extrapolation of the low-temperature experimental data for $\tilde{\kappa}(\tilde{p})$ towards the curve $\tilde{\kappa}= 1/\tilde{p}$ gives $\xi \approx 0.37$, a value that we find consistently for the normalized chemical potential, energy and free energy at our lowest temperatures (see below).

\begin{figure*}
    \centering
    \includegraphics[width=130mm]{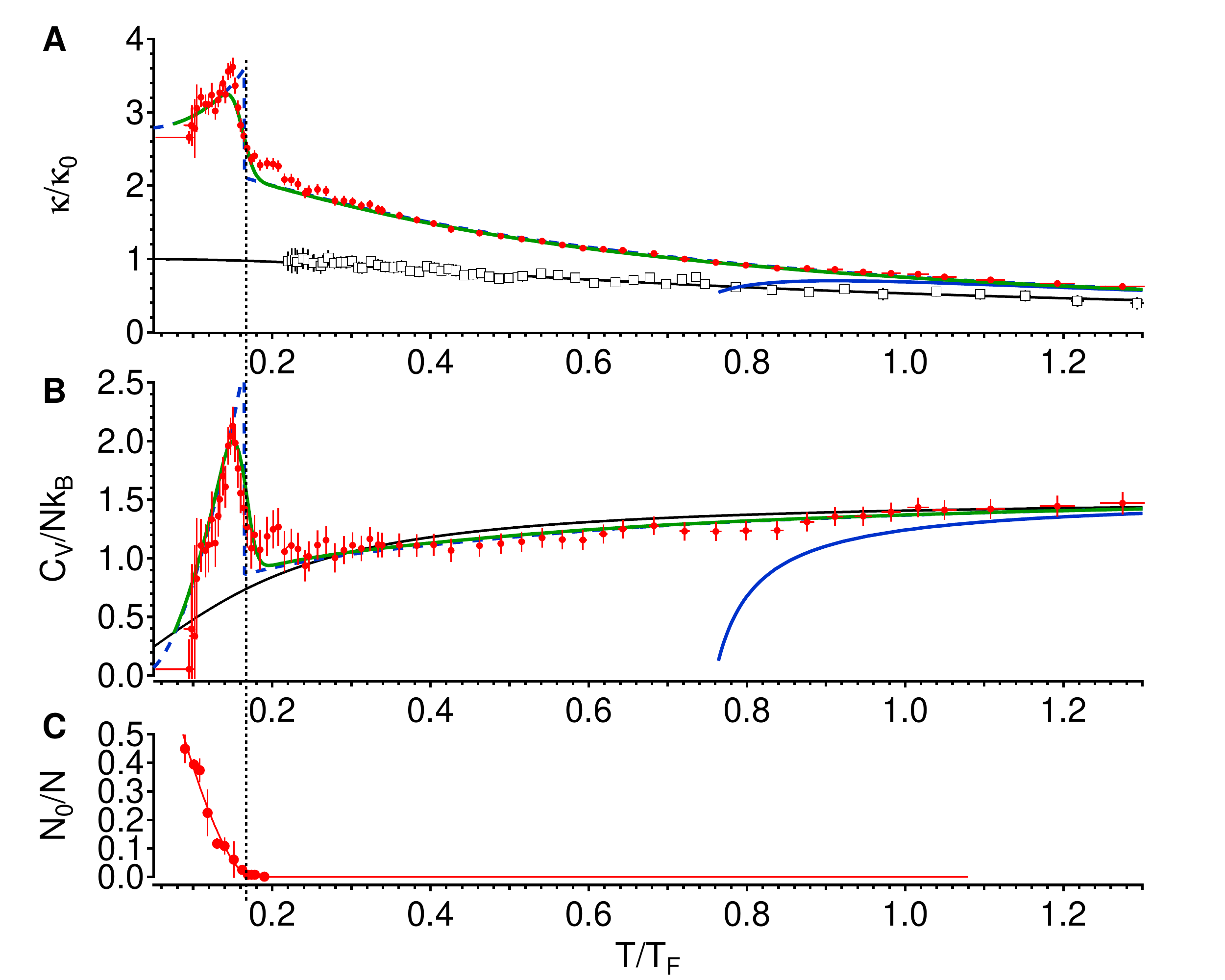}
    \caption{\label{fig2} {\bf A} Normalized compressibility $\tilde{\kappa}=\frac{3}{2}\kappa \,n\,E_F$ and {\bf B} specific heat per particle $C_V/Nk_B$ of a unitary Fermi gas versus reduced temperature $T/T_F$, shown in solid red circles. The black solid curve shows the theory for a non-interacting Fermi gas, the blue solid curve shows the third order Virial expansion for the unitary gas. Black open squares denote data for a non-interacting Fermi gas. The green solid line shows the model from Fig. 1, again including $2 \mu$m imaging resolution. {\bf C} Condensate fraction at unitarity as determined from a rapid ramp to the molecular side of the Feshbach resonance. The onset of condensation coincides with the sudden rise of the specific heat. All error bars show one standard deviation.}
\end{figure*}

From the universal function $\tilde{\kappa}(\tilde{p})$ we obtain {\it all} other thermodynamic quantities of the unitary gas. First of all, it allows translation between the pressure thermometer $\tilde{p}$ and the more familiar normalized temperature $T/T_F$ (where $k_B T_F = E_F$).
Considering that the change in $\tilde{p}$ with $T/T_F$ at constant temperature is related to the change in pressure with density and thus the compressibility, one finds $\frac{{\rm d}\tilde p}{{\rm d}(T/T_F)} = \frac{5}{2} \frac{T_F}{T} \left(\tilde{p}-\frac{1}{\tilde{\kappa}}\right)$ and thus by integration~\cite{ku11supporting}
\begin{equation}
\frac{T}{T_F} = \left (\frac{T}{T_F}\right )_i \exp\left\{\frac{2}{5} \int_{\tilde{p}_i}^{\tilde{p}} {\rm d}\tilde{p} \frac{1}{\tilde{p}-\frac{1}{\tilde{\kappa}}} \right\}.
\end{equation}
Here, $(T/T_F)_i$ is the normalized temperature at an initial normalized pressure $\tilde{p}_i$ that can be chosen to lie in the Virial regime validated above.

Thanks to the relation $E=\frac{3}{2} P \cal{V}$ valid at unitarity~\cite{ho04uni}, we can also directly obtain the heat capacity per particle at constant volume $\cal{V}$~\cite{ku11supporting},
\begin{equation}
  \frac{C_V}{k_B N} \equiv \frac{1}{k_B N} \frac{\partial E}{\partial T}\Big |_{N,\cal{V}} =\frac{3}{5} \frac{{\rm d}\tilde{p}}{{\rm d}\left(T/T_F\right)} = \frac{5}{2}\frac{T_F}{T}\left(\tilde{p} - \frac{1}{\tilde{\kappa}}\right).
\end{equation}

\begin{figure*}[!t]
    \centering
    \includegraphics[width=130mm]{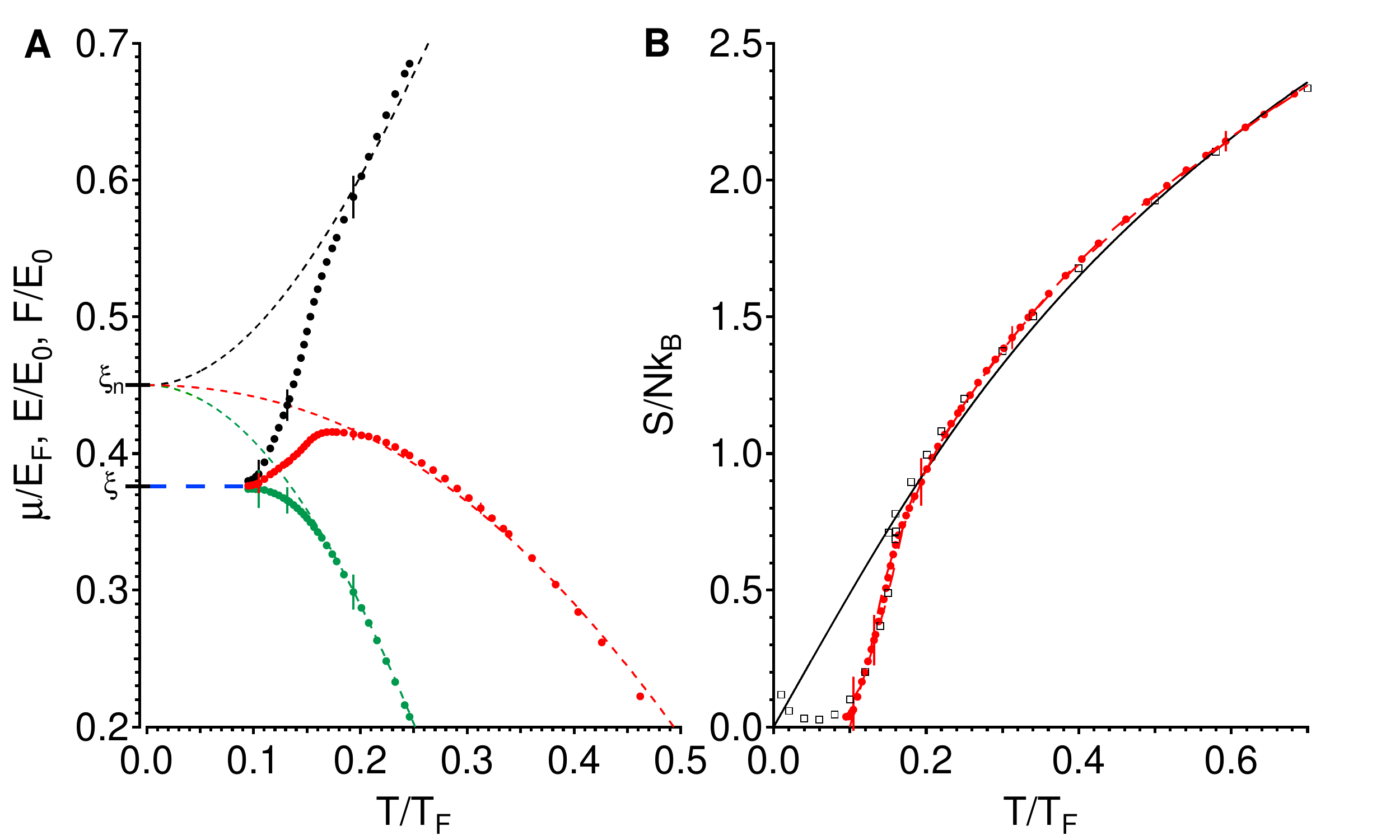}
    \caption{\label{fig3}  {\bf A} Chemical potential $\mu$, energy $E$ and free energy $F$ of the unitary Fermi gas versus $T/T_F$. $\mu$ is normalized by the Fermi energy $E_F$ (red solid circle), $E$ (black solid circle) and $F$ (green solid circle) are normalized by $E_0 = \frac{3}{5} N E_F$. At high temperatures, all quantities approximately track those for a non-interacting Fermi gas, shifted by $\xi_n - 1$ (dashed lines). The peak in the chemical potential signals the onset of superfluidity. In the deeply superfluid regime at low temperatures, $\mu/E_F$, $E/E_0$ and $F/F_0$ all approach $\xi$.
{\bf B} Entropy per particle. At high temperatures, the entropy closely tracks that of a non-interacting Fermi gas (solid line). Below $T/T_F = 0.1$, the entropy per particle reaches values $<0.04\, k_B$.  The open squares are from the self-consistent T-matrix calculation~\cite{haus07bcsbec}. A few representative error bars are shown, representing one standard deviation.}
\end{figure*}

Fig.~2 shows the normalized compressibility and the specific heat as a function of $T/T_F$. At high temperatures, the specific heat approaches that of a non-interacting Fermi gas and eventually $C_V = \frac{3}{2} N k_B$, the value for a Boltzmann gas. A dramatic rise is observed around $T_c/T_F\approx 0.16$, followed by a steep drop at lower temperatures. Such a $\lambda$-shaped feature in the specific heat is characteristic of second order phase transitions, as in the famous $\lambda$-transition in $^4$He~\cite{keesom35}. Jumps in the specific heat are also well-known from superconductors~\cite{tink04sc} and $^3$He~\cite{voll90}. To our knowledge, this is the first time that a specific heat jump has been directly measured in an ultracold atomic gas. Previously, such jumps had only been inferred from derivatives to fit functions that implied a jump~\cite{ensh96,luo09thermo}.
We do not expect to resolve the critical behavior very close to $T_c$, given by $C_V = C_{\rm ns} + A_{+/-}|T-T_c|^{-\alpha}$ with $C_{\rm ns}$ the non-singular part of the specific heat, the critical exponent $\alpha \approx - 0.012$ and amplitudes $A_{+/-}$ above and below the transition~\cite{campostrini06crit}.
In our trapped sample the critical region is confined to a narrow shell due to the spatially varying chemical potential. Based on the estimate in~\cite{poll10crit}, the thickness of the critical shell is 1\% of the cloud size. The finite resolution of our imaging system (2 $\mu$m or about 5\% of the cloud size in the radial direction) suffices to explain the rounding of the singularity expected from criticality. The rounding reduces the jump in the heat capacity at the transition. From our data, we obtain $\Delta C/N\equiv C_s/N-C_n/N$, where $C_s/N$ ($C_n/N$) is the specific heat per particle at the peak (the onset of the sudden rise). Due to the finte resolution, our measurement sets a lower bound on the jump in the specific heat, $\Delta C/C_n\geq 1.0^{+4}_{-1}$. Considering the strong interactions, this is surprisingly close to the BCS value of $1.43$.
Below $T_c$, the specific heat is expected to decrease as $\sim \exp(-\Delta_0/k_B T)$ due to the pairing gap $\Delta_0$. At low temperatures $T \ll T_c$ the phonon contribution $\propto T^3$ dominates~\cite{haus07bcsbec}. This behavior is consistent with our data, but the phonon regime is not resolved.

To validate our in-situ measurements of the superfluid phase transition, we have employed the rapid ramp method to detect fermion pair condensation~\cite{rega04,zwie04rescond}. The results are shown in Fig. 2{\bf C}, and demonstrate that the onset of condensation and the sudden rise in specific heat and compressibility all occur at the same critical temperature, within the error bars. Previous experimental determinations of $T_c/T_F$ of the homogeneous unitary Fermi gas relied on a temperature calibration that did not agree with the Virial expansion~\cite{hori10thermo}, or obtained $T_F$ from the $\mu$-derivative of a pressure fit, and $T_c$ from a construction implying a first-order transition~\cite{nasc10thermo}. Here we determine $T_c/T_F$ directly from the density profiles, finding a sudden rise in the specific heat and the onset of condensation at $T_c/T_F = 0.167(13)$. This value is determined as the midpoint of the sudden rise, and the error is assessed as the shift due to the uncertainty of the Feshbach resonance~\cite{ku11supporting}. This is in very good agreement with theoretical determinations, for example, the self-consistent T-matrix approach that gives $T_c/T_F \approx 0.16$~\cite{haus94,haus07bcsbec}, and Monte-Carlo calculations which give $T_c/T_F = 0.173(6)$~\cite{goul10tc} and $0.152(7)$~\cite{buro06TC}. It disagrees with $T_c/T_F = 0.23(2)$~\cite{bulg06TC}, but is close to a later determination by the same group of $T_c/T_F \lesssim 0.15(1)$~\cite{bulg08qmc}. There is a current debate on the possibility of a pseudo-gap phase of preformed pairs above $T_c$~\cite{gaeb10pseudo,nasc10thermo}. A pairing gap for single-particle excitations above the transition should be signaled by a suppression of the specific heat already above $T_c$, which is not observed in our measurements.

\begin{figure*}[!t]
    \centering
    \includegraphics[width=130mm]{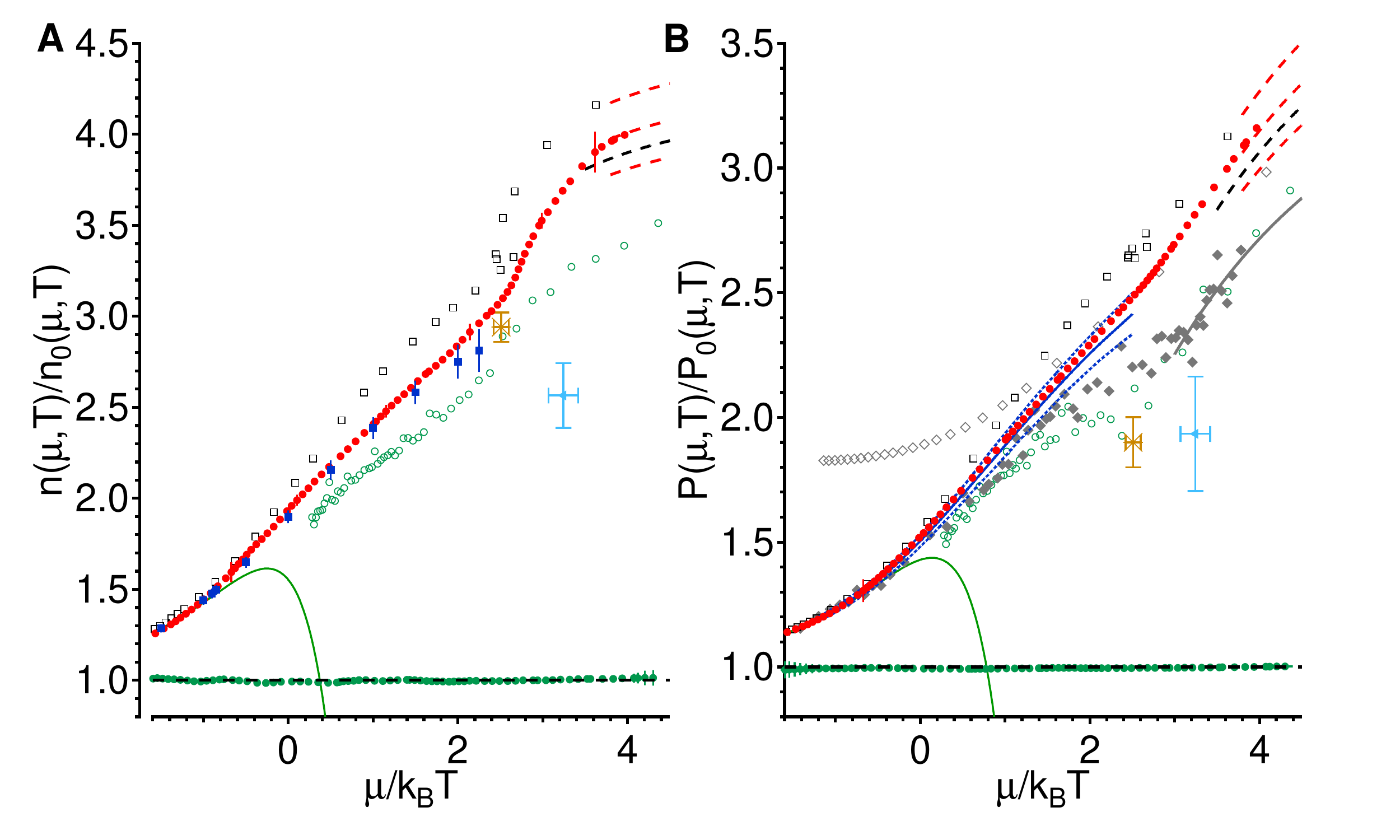}
    \caption{\label{fig4}  {\bf A} Density and {\bf B} pressure of a unitary Fermi gas versus $\mu/k_B T$, normalized by the density and pressure of a non-interacting Fermi gas at the same chemical potential $\mu$ and temperature $T$. Red solid circles: experimental EoS. Dashed lines: low-temperature behavior with $\xi = 0.364, 0.376$ and $0.389$. Black dashed line: low-temperature behavior from the $\xi$ upper bound $\xi = 0.383$~\cite{forb10xibound}. Green open circles and black dashed line at 1.0: MIT experimental density and pressure, and theory for the ideal Fermi gas. Blue solid squares (blue band): Diagrammatic Monte Carlo~\cite{vanhouke11crossval} for density (pressure). Solid green line: 3rd order Virial expansion. Open black squares: self-consistent T-matrix~\cite{haus07bcsbec}. Open green circles:~\cite{bulg06TC}. Orange star:~\cite{goul10tc}. Blue star:~\cite{buro06TC}. Solid diamonds: ENS experiment~\cite{nasc10thermo}. Open diamonds: Tokyo experiment~\cite{hori10thermo}.}
\end{figure*}

From the definition of the compressibility $\kappa = \frac{1}{n^2}\frac{\partial n}{\partial\mu}|_T$ we can obtain the reduced chemical potential $\mu/E_F$ as a function of the reduced temperature, see Figure 3{\bf A} and~\cite{ku11supporting}. Here, for the first time, this function is obtained from measured quantities, rather than from numerical derivatives of data that involved uncontrolled thermometry~\cite{hori10thermo}. Around $T/T_F\sim 0.25$ to $1$, the chemical potential is close to that of a non-interacting Fermi gas, shifted by $(\xi_n-1)E_F$ due to interactions present already in the normal state, with $\xi_n \approx 0.45$. However, unlike a normal Fermi gas, the chemical potential attains a maximum of $\mu/E_F = 0.42(1)$ at $T/T_F = 0.171(10)$, and then decreases at lower temperatures. This is expected for a superfluid of paired fermions~\cite{haus07bcsbec}. As the temperature is increased from zero in a superfluid, phonons (sound excitations) emerge that increase the chemical potential $\mu$. In addition, fermion pairs start to break and single fermions contribute increasingly to the chemical potential with increasing temperature.
At $T_c$, $\mu/E_F$ must have a sharp change in slope, as ${\rm d}(\mu/E_F)/{\rm d}(T/T_F)$ involves the singular compressibility. Indeed, the self-consistent T-Matrix calculation shows a very clear peak in $\mu/E_F$ near $T_c$~\cite{haus07bcsbec}, in agreement with our observation.
At low temperatures, the reduced chemical potential $\mu/E_F$ saturates to the universal value $\xi$. As the internal energy $E$ and the free energy $F$ satisfy $E(T)>E(T=0) = \frac{3}{5} N \xi E_F = F(T=0)>F(T)$ for all $T$, the reduced quantities $f_E \equiv \frac{5}{3}\frac{E}{N E_F}=\tilde{p}$ and $f_F \equiv \frac{5}{3}\frac{F}{N E_F} = \frac{5}{3}\frac{\mu}{E_F} - \frac{2}{3}\tilde{p}$ provide upper and lower bounds for $\xi$~\cite{cast11unitary}, shown in Fig. 3{\bf A}. Taking the coldest points of these three curves and including the systematic error due to the effective interaction range, we find $\xi=0.376(5)$. The uncertainty in the Feshbach resonance is expected to shift $\xi$ by at most $2\%$~\cite{ku11supporting}.
This value is consistent with a recent upper bound $\xi < 0.383$~\cite{forb10xibound}, is close to $\xi=0.36(1)$ from a self-consistent T-matrix calculation~\cite{haus07bcsbec}, and agrees with $\xi=0.367(9)$ from an epsilon expansion~\cite{arno07xi}. It lies below earlier estimates $\xi=0.44(2)$~\cite{carl03} and $\xi=0.42(1)$~\cite{astr04} via fixed-node quantum Monte-Carlo that provide upper bounds on $\xi$. Our measurement agrees with several less accurate experimental determinations~\cite{gior08review}, but disagrees with the most recent experimental value $0.415(10)$ that was used to calibrate the pressure in~\cite{nasc10thermo}, shown in Fig. 4{\bf B}.

From the energy, pressure and chemical potential, we can obtain the entropy $S = \frac{1}{T}(E + PV - \mu N)$. Shown in Figure 3{\bf B} is the entropy per particle $S/Nk_B=\frac{T_F}{T}(\tilde{p}-\frac{\mu}{E_F})$ as a function of $T/T_F$. At high temperatures, $S$ is close to the entropy of an ideal Fermi gas at the same $T/T_F$. Down to $T_c$, neither the non-interacting nor the unitary Fermi gas has $S/N\ll k_B$. Also, the specific heat $C_V$ is not linear in $T$. Thus it is questionable to identify the normal regime as a Landau Fermi Liquid, although some thermodynamic quantities agree surprisingly well with the expectation for a Fermi liquid (see~\cite{nasc10thermo} and~\cite{ku11supporting}). Below about $T/T_F = 0.17$ the entropy starts to strongly fall off compared to that of a non-interacting Fermi gas, which we again interpret as the freezing out of single-particle excitations due to formation of fermion pairs. Far below the critical temperature for superfluidity, phonons dominate. They only have a minute contribution to the entropy~\cite{haus07bcsbec}, less than 0.02 $k_B$ at $T/T_F = 0.1$, consistent with our measurements. At the critical point we obtain $S_c = 0.73(13) Nk_B$, in agreement with~\cite{haus07bcsbec}. It is encouraging for cold atom experiments that we obtain very low entropies, less than $0.04\,N k_B$, far below critical entropies required to reach magnetically ordered phases of fermions in optical lattices.

From the chemical potential $\mu/E_F$ and $T/T_F = \frac{4\pi}{(3\pi^2)^{2/3}}\frac{1}{(n\lambda^3)^{2/3}}$, we finally obtain the density EoS $n(\mu,T) \equiv \frac{1}{\lambda^3} f_n(\beta\mu)$, with the de Broglie wavelength $\lambda=\sqrt{\frac{2\pi\hbar^2}{mk_BT}}$. The pressure EoS follows as $P(\mu,T) \equiv \frac{k_B T}{\lambda^3}f_P(\beta\mu)$ with $f_P = \frac{2}{5}\frac{T_F}{T}\tilde{p} f_n(\beta\mu)$.
Fig. 4 shows the density and pressure normalized by their non-interacting counterparts at the same chemical potential and temperature. For the normal state, a concurrent theoretical calculation employing a new Monte-Carlo method agrees excellently with our data~\cite{vanhouke11crossval}. Our data deviates from a previous experimental determination of the pressure EoS~\cite{nasc10thermo} that was calibrated with an independently measured value of $\xi = 0.415(10)$~\cite{nasc10thesis} - a value which agrees neither with our value nor a recent theoretical upper bound~\cite{forb10xibound}. Our data also disagrees with the energy measurement in~\cite{hori10thermo} that used a thermometry inconsistent with the Virial expansion~\cite{luo09thermo}. Around the critical point the density shows a strong variation, while the pressure, the integral of the density over $\mu$ at constant $T$, is naturally less sensitive to the superfluid transition.

In conclusion, we have performed a high-precision experimental study of the thermodynamics of a unitary Fermi gas across the superfluid transition. For the first time, we observe direct signatures of the superfluid phase transition in various thermodynamic quantities, such as a sudden rise in the compressibility and the specific heat. The critical temperature for superfluidity at unitarity is found to be $T_c = 0.167(13)T_F$, the critical entropy is $0.73(13)Nk_B$, and we set a lower bound on the jump in the specific heat, $\Delta C/C_n \geq 1.0^{+4}_{-1}$. We also find a new experimental value for the Bertsch parameter $\xi = 0.376(5)$, subject to at most a $2\%$ shift due to the uncertainty in the Feshbach resonance position. We have demonstrated that precision many-body determinations of phase transitions involving fermionic atoms are possible at the few percent level and without any fits or input from theory, enabling validation of theories for strongly interacting matter. Similar unbiased methods can be applied to other systems, for example, two-dimensional Bose and Fermi gases or fermions in optical lattices.

\begin{acknowledgments}
We would like to thank B. Svistunov, N. Prokof'ev, F. Werner for fruitful discussions, Z. Hadzibabic for a critical reading of the manuscript, the authors of~\cite{buro06TC,goul10tc,bulg06TC,haus07bcsbec,hori10thermo,nasc10thermo} for kindly providing us with their data,
 and Andr\'e Schirotzek for help during the early stages of the experiment. This work was supported by the NSF, AFOSR-MURI, ARO-MURI, ONR, DARPA YFA, a grant from the Army Research Office with funding from the DARPA OLE program, an AFOSR PECASE, the David and Lucille Packard Foundation, and the Alfred P. Sloan Foundation.
\end{acknowledgments}


%

\section*{Supplemental Materials and Methods}
\setcounter{figure}{0}
\makeatletter 
    \renewcommand{\thefigure}{S\@arabic\c@figure} 
\makeatother

\subsection*{Obtaining density versus potential curves from absorption images of trapped samples}

In this section we provide details on the determination of the density from in-situ images of trapped samples. The experimental set up is described in~\cite{varenna08supp}. Fermionic $^6$Li is cooled to degeneracy via sympathetic cooling with $^{23}$Na. A two-state mixture of the two lowest hyperfine states of $^6$Li is prepared and brought into the strongly-interacting regime at a broad Feshbach resonance at $834.15$ G in a hybrid magnetic and optical trap, where the mixture is further cooled. We measure the column density of the gas via high-resolution {\it in situ} absorption imaging.

The trapping potential is provided by a magnetic field saddle potential that is confining along the $z$-direction, and a gaussian laser beam (wavelength $1064$ nm, waist $w = 120\; \mu$m) propagating along $z$. Due to the large waist and correspondingly large Rayleigh range of the gaussian beam, the trapping potential is to an excellent approximation harmonic in the $z$-direction, with a measured trapping frequency of $\omega_z = 2\pi \cdot 22.83(5) \;\rm Hz$. Waist $w$ and power $\cal{P}$ of the laser beam determine the trapping potential along the radial direction. The total potential can be excellently modeled as
\begin{eqnarray}
\nonumber V(\rho,z) &=& V_\rho(\rho) + V_z(z)\\
 &\equiv& \frac{1}{2} m \omega_z^2 z^2 -\frac{1}{4}m \omega_z^2 \rho^2 - \alpha \frac{\cal{P}}{\frac{\pi}{2} w^2} e^{-2 \rho^2/w^2},
\end{eqnarray} where $\alpha$ is the polarizability. We can safely neglect in this model the 0.01\% contribution to the axial trapping frequency from the variation of the laser power along the $z$-direction.
We do not rely on the waist and power measurement of the laser beam but in fact measure the trapping potential directly using the atomic distribution itself, as will be outlined below.

To obtain the 3D density from the measured optical density $\mathrm{OD}(x,z) = \sigma n_{2D}(x,z)$ with effective absorption cross section $\sigma$, we employ the inverse Abel transform for each $z$~\cite{shin06phasesupp}:
$$n_{3D}(x,z) = -\frac{1}{\pi} \int_x^{\infty} {\rm d}\rho\; \frac{1}{\sqrt{\rho^2-x^2}} \frac{\partial n_{2D}}{\partial \rho}(\rho,z)$$
Cylindrical symmetry, i.e. the equality of the waist of the optical potential along the transverse $x$ and $y$ direction, is measured via a beam profiler to be better than 1\% and confirmed via the observation of long-lived vortices~\cite{zwie05vortsupp}. Cylindrical symmetry is not required: As long as equipotential lines in the $(x,y)$ plane transverse to the $z$-direction can be parameterized as an ellipse $x^2/a^2 + y^2/b^2 = {\rm const}$, the inverse Abel formula for the 3D density would be modified only through an overall factor $b/a$:
$$n_{3D}(x,z) = -\frac{a}{b}\frac{1}{\pi} \int_{x}^{\infty} {\rm d}\rho\; \frac{1}{\sqrt{\rho^2-x^2}} \frac{\partial n_{2D}}{\partial \rho}(\rho,z)$$
The absolute scale of density is calibrated through normalization via the measured effective absorption cross section of a non-interacting Fermi gas, so this prefactor, even if different from unity, is simply absorbed in the effective absorption cross section. Absorption imaging is performed at low intensity $I/I_{\rm SAT} < 0.07$ to avoid saturation, and the total number of photons scattered is kept small (around 4) by using a short pulse duration to avoid optical pumping and an increasing Doppler shift~\cite{sanner10supp}. Simulations have confirmed that non-linear effects in our imaging procedure distort the density profile by much less than 1\%.

To obtain a direct measurement of the trapping potential, one can employ the local density approximation: equidensity lines must be equipotential lines, and the potential is excellently known in the axial ($z$) direction, thus calibrating the potential everywhere. For this measurement, one may average many independent 3D density profiles, even if the atom number or temperature vary from shot to shot. The averaged density along $z$, $n_{3D}^{\rm avg.}(0,z) = n_{3D}^{\rm avg.}(V(0,z))$ can be inverted to find $V(n_{3D})$, so that the density at any point $(\rho,z)$ in the trap gives $V(\rho,z)$. The resulting potential can be fit with our model, and results in a waist that agrees with the measured waist to within one percent, and a power that agrees with the measured power to within 10\% (a typical error for standard power meters).
Thus anharmonicities in the trapping potential have been incorporated into the experimental analysis of density profiles exactly. Previous measurements attempted to characterize anharmonic traps via measured trapping frequencies~\cite{hori10supp,nasc09supp}. However, anharmonicities cause such frequencies to depend on the trap filling, rendering that method less reliable.
Knowing the trapping potential, a single experimental density profile $n_{3D}(\rho,z)$ can now be averaged over equipotential lines $V(\rho,z) = V_0$, yielding a low-noise determination of the density as a function of the local potential $V_0$.

The local density approximation (LDA) is well fulfilled as long as the cloud size (typically $\sim 40\,\rm \mu m$ in the narrow, radial direction) is large compared to the characteristic length of spatial correlations in the gas, on the order of $\sim 1/k_F = 400\,\rm nm$~\cite{poll10critsupp}. Violations of LDA due to criticality close to the superfluid transition have not been detected, likely due to our finite optical resolution (about $2-3$ $\mu$m).

The profiles $n(V)$ for varying atom numbers, trap parameters and degree of evaporative cooling are then used to calculate the normalized compressibility and the pressure, as described in the text.

\subsection*{Compressibility equation of state and relation to the specific heat and reduced temperature}

The compressibility $\kappa$, the pressure $P$ and the density $n$ are related by an equation of state $\kappa(n,P)$. At unitarity, all these thermodynamic quantities are related to universal functions of $\beta\mu$.
Let us define $P=\frac{1}{\beta\lambda^3}f_P(X)$, then $n = \frac{1}{\lambda^3} f_P'(X)\equiv\frac{1}{\lambda^3}f_n(X)$, and $\kappa n^2 = \frac{\beta}{\lambda^3} f_P''(X)$,
with $X = \beta\mu$.

The experiment determines the density as a function of the local potential, $n(V)$. We have $\kappa n^2 = \frac{\partial n}{\partial \mu} = -\frac{\partial n}{\partial V}$.
We normalize $\kappa$ by the compressibility of a non-interacting, zero-temperature Fermi gas $\kappa_0 = \frac{3}{2} \frac{1}{n \epsilon_F}$.
We find $\tilde{\kappa} \equiv \frac{\kappa}{\kappa_0} = \frac{\partial E_F}{\partial \mu} =  - \frac{\partial E_F}{\partial V}$, where $E_F=\frac{\hbar^2(3\pi^2 n)^{2/3}}{2m}$ is the Fermi energy, with $m$ being the mass of the atom.

At unitarity, $\tilde{\kappa}$ is a universal function of $T/T_F$ or, equivalently, of $n \lambda^3$.
If we replace the temperature by the pressure $P$, another dimensionless quantity that we can form is $\tilde{p} = \frac{P}{\frac{2}{5} n \epsilon_F}$, which is the pressure normalized by the zero-temperature limit for a non-interacting gas. One can trade the unknown thermometer $T/T_F$ for the known, new thermometer $\tilde{p}$.
For every profile, one can determine $\tilde{\kappa}(V)=-\frac{dE_F(V)}{dV}$ and $\tilde{p}(V)=\frac{\int_{V}^{\infty}dV' n(V')}{\frac{2}{5}n(V)E_F(V)}$, and then plot $\tilde{\kappa}$ vs $\tilde{p}$. Even though obtaining $\tilde{\kappa}$ involves taking a derivative of the data, accumulation of data reduces the noise. Data from all profiles within a bin of $\tilde{p}$ is averaged to give the EoS. The statistical error bar is taken to be the standard error.

In terms of the function $f_P$ we have $\tilde{\kappa} = \frac{2}{3} \frac{T_F}{T} \frac{f_P''}{f_P'}$ and $\tilde{p} = \frac{5}{2} \frac{T}{T_F} \frac{f_P}{f_P'}$. The different thermometers, $\tilde{p}$, $X = \beta\mu$ and $T/T_F$ depend on each other in the following way:
\begin{eqnarray}
\frac{{\rm d}\tilde p}{{\rm d}X}&=&\frac{5}{2} \frac{T}{T_F}(1 - \tilde{\kappa}\tilde{p})\\
\frac{{\rm d}(T/T_F)}{{\rm d}X}&=& -\tilde{\kappa}\left(\frac{T}{T_F}\right)^2\\
\frac{{\rm d}\tilde p}{{\rm d}(T/T_F)}&=& \frac{5}{2} \frac{T_F}{T} \left(\tilde{p}-\frac{1}{\tilde{\kappa}}\right).
\end{eqnarray}
Note that apart from a prefactor, the last equation is just the heat capacity per particle (using $E=\frac{3}{2}P\mathcal{V}$ valid at unitarity):
\begin{equation}
\frac{C_V}{k_B N} = \frac{1}{k_B N}\frac{{\rm d}E}{{\rm d}T}\Big |_{N,\mathcal{V}} = \frac{3}{5} \frac{{\rm d}\tilde{p}}{{\rm d}\left(T/T_F\right)} =  \frac{3}{2} \frac{T_F}{T} \left(\tilde{p}-\frac{1}{\tilde{\kappa}}\right).
\end{equation}
Further simplifying, we have $\frac{{\rm d}\tilde p}{{\rm d}\ln(T/T_F)}=\frac{5}{2}\left(\tilde{p}-\frac{1}{\tilde{\kappa}}\right)$ which can be integrated to give
\begin{equation}
\frac{T}{T_F} = \left (\frac{T}{T_F}\right )_i \exp\left\{\frac{2}{5} \int_{\tilde{p}_i}^{\tilde{p}} {\rm d}\tilde{p} \frac{1}{\tilde{p}-\frac{1}{\tilde{\kappa}}} \right\},
\label{eq:tovertf}
\end{equation}
where $\tilde{\kappa}$ is known as a function of $\tilde{p}$, and $(T/T_F)_i$ is the temperature at the initial normalized pressure $\tilde{p}_i$.
This function relates the thermometer $T/T_F$ to the pressure thermometer $\tilde{p}$.

There are several ways to obtain $\beta\mu$ vs $T/T_F$. One can consider $\tilde{\kappa}$ as a function of $T/T_F$ and obtain
\begin{equation}
\beta\mu= (\beta\mu)_i - \int_{T_i/T_F}^{T/T_F} {\rm d}({\textstyle \frac{T}{T_F}})  \frac{1}{\tilde{\kappa}} \left(\frac{T_F}{T}\right)^2,
\end{equation}
or one can consider $T/T_F$ as a function of $\tilde{p}$ and obtain
\begin{equation}
\beta\mu = (\beta\mu)_i + \frac{2}{5} \int_{\tilde{p}_i}^{\tilde{p}} {\rm d}\tilde{p} \,\frac{T_F}{T} \frac{1}{1 - \tilde{\kappa}\tilde{p}}.
\end{equation}

Together with $T/T_F = \frac{4\pi}{(3\pi^2)^{2/3}}\frac{1}{f_P'^{2/3}}$, this now gives $f_P'(\beta\mu)=f_n(\beta\mu)$. We then get directly $f_P = \frac{2}{5} \tilde{p} \frac{T_F}{T} f_P'$ and $f_P'' = \frac{3}{2}\frac{T}{T_F} f_P' \tilde{\kappa}$. In addition, for the entropy per particle we have $S/N k_B = \frac{T_F}{T}(\tilde{p}-\frac{\mu}{E_F})$.

We validate the method via the measurement of the equation of state of a non-interacting Fermi gas. This provides a check for the effective absorption cross section: If the cross section had been determined too large by a factor of $\gamma$, $\tilde{\kappa} \propto \gamma^{-2/3}$ would be too small, and $\tilde{p} \propto \gamma^{2/3}$ too large. Note, however, that the determination of $T/T_F$ from $\tilde{\kappa}(\tilde{p})$ is insensitive to errors in the absorption cross section, as all factors of $\gamma$ cancel in Eq.~\ref{eq:tovertf}.
Systematic errors that are unique to the unitary gas are discussed in the following.

\subsection*{Systematic errors from non-universal scattering properties}

In the following we discuss systematic errors in our experiment from possible non-universal behavior.
The $s$-wave scattering amplitude $f(k)$ for atom-atom collisions can be written as
$$f = \frac{1}{u(k) - ik}$$
where $u(k) = -\frac{1}{a} + \frac{1}{2} r_e k^2 + \dots$, $a$ is the scattering length, and $r_e$ the effective range. The latter can generally differ from the range of the interatomic potential, $b$. To be in the universal regime at unitarity where the universal relation $n\lambda^3 = f(\beta\mu)$ holds, we require that $k_F a \gg 1$, $k_F r_e \ll 1$, $k_F b \ll 1$~\cite{wern09closedsupp}.
Otherwise, the additional length scales $a$, $r_e$ and $b$ will feature in the more complex equation of state that would be written as $n\lambda^3 = f(\beta\mu,\lambda/a, \lambda/r_e,\lambda/b)$.
We discuss the possible influence of the various terms separately in the following.

\subsubsection{Effect of the uncertainty in the Feshbach resonance position}

The Feshbach resonance between the lowest two hyperfine states in $^6$Li lies at $(834.15 \pm 1.5)\,\rm G$~\cite{bart04feshsupp}.
The systematic error from the uncertainty in the Feshbach resonance position of $\Delta B = 1.5 \,\rm G$ can be estimated using the contact~\cite{tan08energysupp}. As the scattering length varies, the change in the energy of the system is given by
$$\left(\frac{\partial E}{\partial a^{-1}}\right)_{S,N,\mathcal{V}} = \frac{\hbar^2}{4\pi m} \cal{C} V$$
where $\cal{C}$ is the contact density. From this relation, the change in the pressure and, via $n = \frac{\partial P}{\partial \mu}|_{T}$, the change in density with $a$ can be derived. Diagrammatic Monte-Carlo calculations yield the temperature dependence of the contact~\cite{vanhouke11supp}.
We can thus predict the true density profiles starting with our measured normalized densities. The result is shown as the triangles pointing up and down in Fig. S1.

\begin{figure}
    \centering
    \includegraphics[width=84mm]{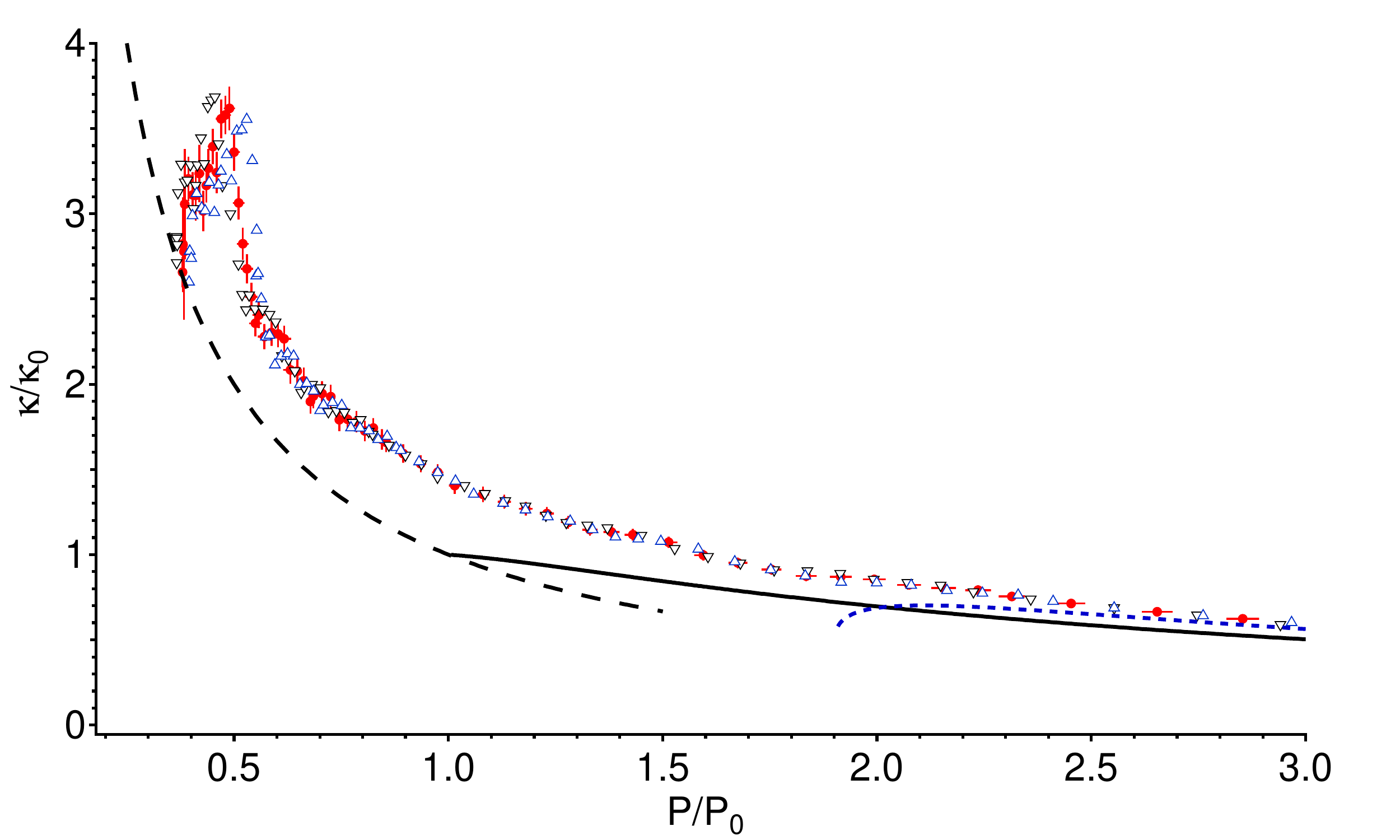}
    \caption{\label{figs1} The effect of uncertainty in the Feshbach resonance position on the equation of state $\tilde{\kappa}$ versus $\tilde{p}$. The experimental result at $834.15$ G is shown in red solid circles. This would be the EoS of the unitary Fermi gas if the Feshbach resonance is located at $B_0=834.15$ G. If the true location of the resonance is at $\pm 1.5$ G away, one can estimate the true EoS for the unitary Fermi gas using the contact (see text). The result is shown as triangles pointing up (down) if the true resonance position is at $832.65$ G ($835.65$ G). Also shown are the third order virial expansion (blue solid curve), the non-interacting EoS (black solid curve), and the curve $\tilde{\kappa}=1/\tilde{p}$ on which the curve $\tilde{\kappa}$ versus $\tilde{p}$ must end.}
\end{figure}

From these upper and lower bounds on the true equation of state we can deduce upper and lower bounds for the normalized chemical potential, energy and free energy, pressure and entropy, which result from different combinations of density and pressure.
All errors grow as the temperature is lowered, as the contact increases. At the lowest temperatures, the error in the density is $3\%$, in the pressure $1\%$, in the chemical potential (and therefore $\xi$) $2\%$, in the normalized energy (free energy) $4\%$ ($1\%$).
As it turns out, the entropy per particle $S/N$ is only very weakly sensitive to the uncertainty in the resonance position, as the systematic error in the pressure and density cancel to a high degree. At the lowest temperatures, the error has grown to only $\pm 0.08 k_B$. At higher temperatures, the systematic error in $S/N$ from the error in $B_0$ is much smaller than the statistical error bars.
Note that $\Delta B$ is much larger than the variation of the magnetic field along the axial direction of the cloud of about $15 \,\rm mG$ due to the magnetic confinement, which contributes only a 0.01\% error in $\xi$.

\subsubsection{Effective range correction}

The effective range near a Feshbach resonance at scattering length $a$ can be modeled as~\cite{wern09closedsupp}:
$$r_e = -2 R_*(1-\frac{a_{\rm bg}}{a})^2 + \frac{4 b}{\sqrt{\pi}} - \frac{2 b^2}{a}$$
where $R_* = 0.0269\,\rm nm$ and $b = 2.1 \,\rm nm$ for $^6$Li~\cite{wern09closedsupp}. At our typical densities, $1/k_F \approx 400 \,\rm nm$ and thus $k_F R_* < 10^{-4}$. As this parameter is very small, errors due to spatial variation of the scattering length, mediated by the optical trap, are vanishingly small~\cite{wern09closedsupp}. As $a\gg b$ across the Feshbach resonance, the effective range is reduced to a constant $r_e = 4.7\,\rm nm$ for essentially all magnetic fields $B \gtrsim 650 \,\rm G$~\cite{naid10efimovsupp}.

At our highest densities at low temperatures, $k_F r_e = 0.012$. All thermodynamic potentials and thus also the ground state energy $E = \frac{3}{5} \xi N E_F$ are expected to depend linearly on this parameter~\cite{wern10exactsupp}. From the Quantum Monte-Carlo values of $\xi$ versus effective range given in~\cite{gand11becbcssupp} one can deduce a linear behavior of the upper bound on $\xi$:
$$\xi(k_F r_e) = \xi(r_e=0) + 0.26(6) \cdot k_F r_e + \dots$$
With this dependence of $\xi$ on the effective range, the error on $\xi$ in our experiment is at most +0.003 or +0.8\%. This is small compared to the error due to the uncertainty in the Feshbach resonance position.

\begin{figure}
    \centering
    \includegraphics[width=84mm]{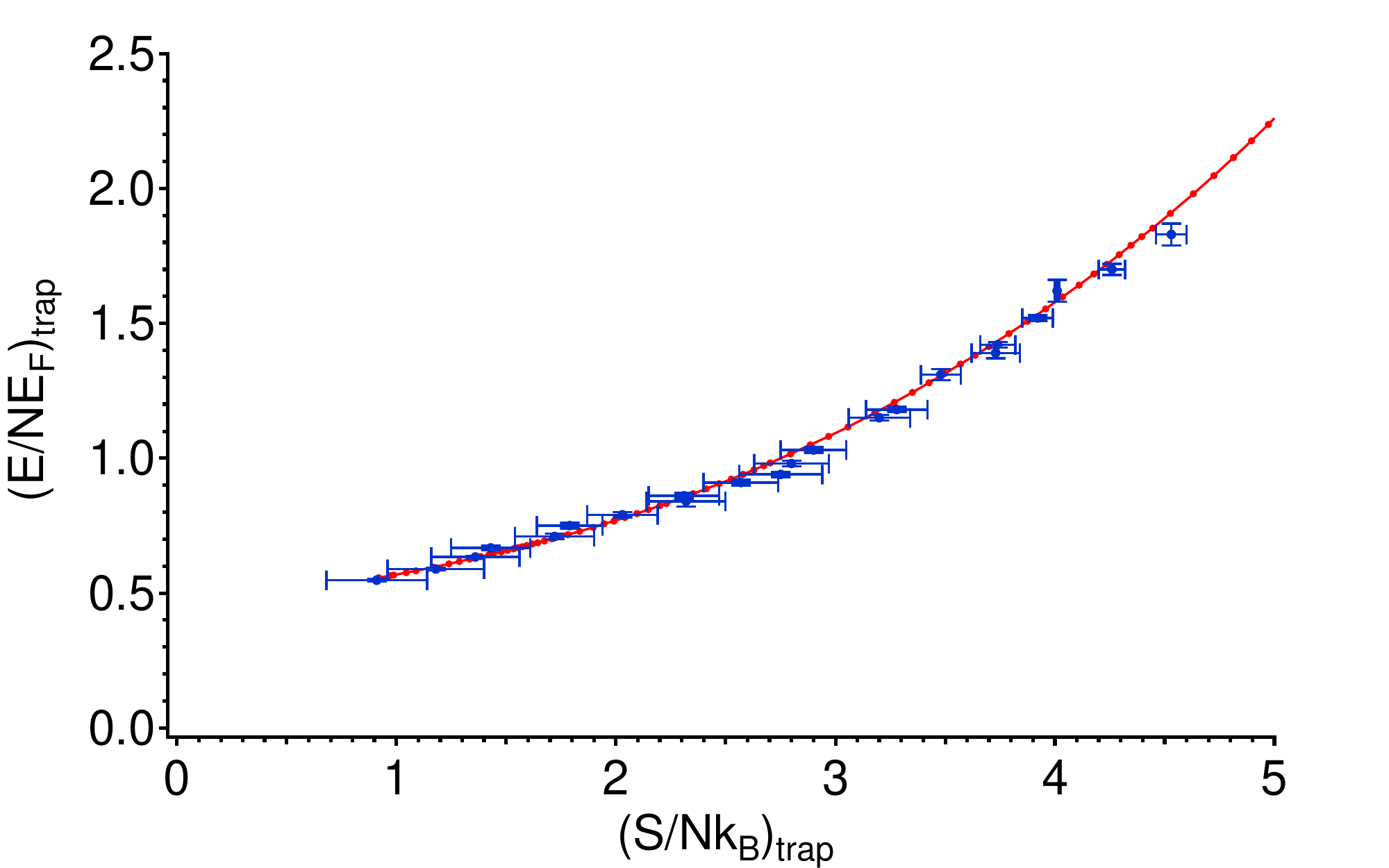}
    \caption{\label{figs2} Energy vs entropy of a harmonically trapped gas at unitarity, deduced from our measured bulk equation of state (red solid dots). The curve is smooth, as required by thermodynamics. Blue diamonds: Duke experiment~\cite{luo07entropysupp}, for which an interaction correction has been applied to the entropy.}
\end{figure}

\subsection*{Energy vs entropy of the harmonically trapped gas}

In Fig. S2 we show the energy versus entropy for a harmonically trapped Fermi gas at unitarity, determined from our experimental determination of the homogeneous density EoS and temperature. The data are compared with the earlier measurement of the same quantities by the Duke group~\cite{luo07entropysupp}, after applying a correction of the entropy for the finite interaction strength in that measurement. The agreement is excellent. Note that the energy vs entropy curve is smooth and on its own does not allow for a determination of the critical energy or entropy, in contrast to earlier interpretations of the Duke data~\cite{luo07entropysupp}.
From the bulk thermometry, we obtain $E_{c,{\rm trap}} = 0.698(23)N E_{F,{\rm trap}}$ and $S_{c,{\rm trap}} = 1.70(10) N k_B$.

\subsection*{Discussion on possible Fermi Liquid behavior}

In Fig. S3 and S4 we plot the density and pressure, normalized by that of a non-interacting Fermi gas at zero temperature, versus $(T/\mu)^2$. The linear behavior of the normalized density and pressure at high temperatures is reminiscent of a Fermi liquid.
However, in the range shown, $T/\mu$ is not a small parameter. The superfluid transition leads to a dramatic upturn at low $T$ in the normalized density at low temperatures. This is directly related to the downturn in the chemical potential seen in Fig. 3 {\bf A}, as $n(\mu,T)/n_0(\mu,0) \propto (\mu/E_F)^{-3/2}$. The minimum value of the normalized density occurs at $T/\mu = 0.41(5)$, close to $T_c/\mu_c = 0.40(3)$, where $T_c$ is determined from the midpoint in the specific heat jump and validated via the condensate fraction measurement.

\begin{figure}[h]
    \centering
    \includegraphics[width=84mm]{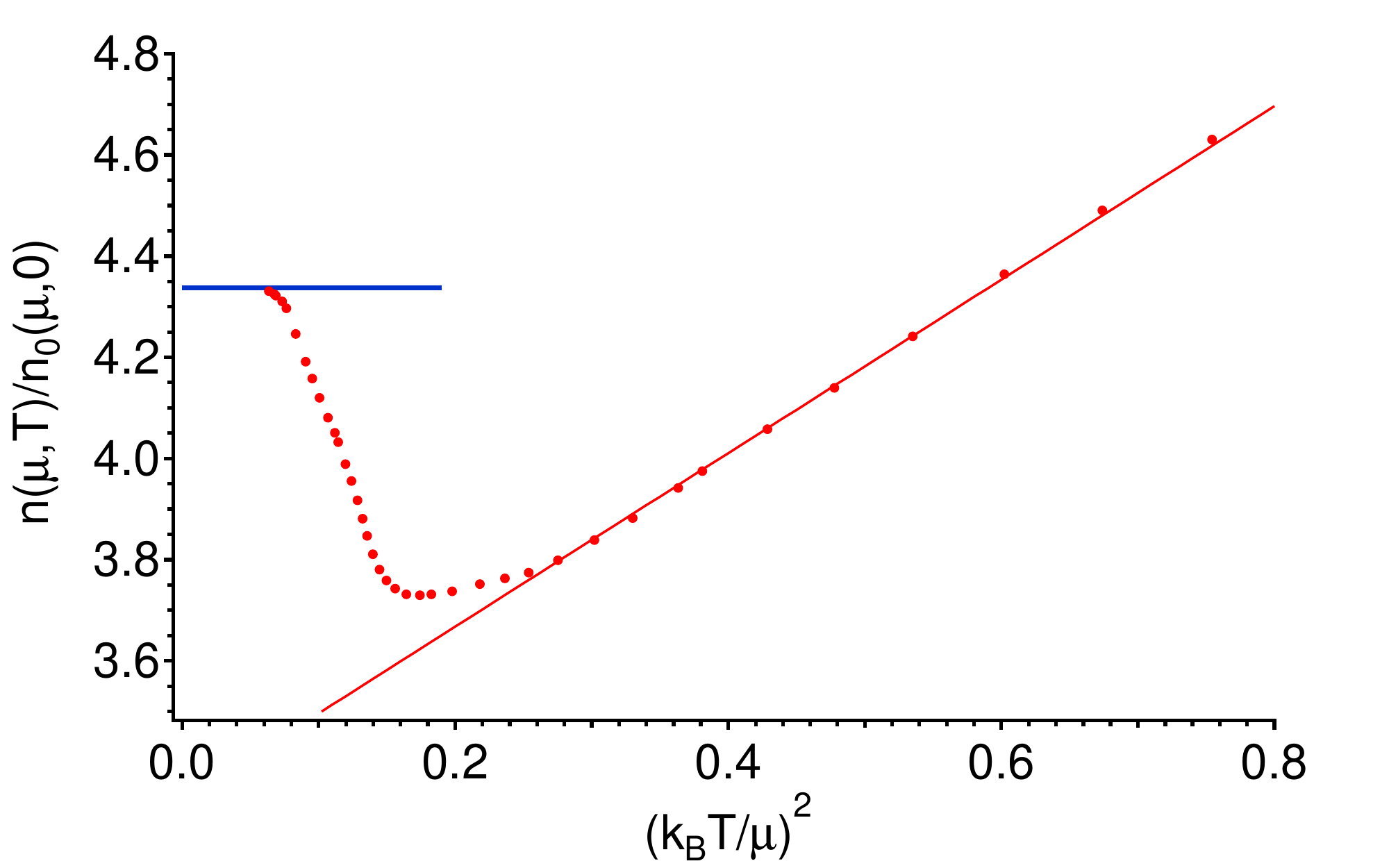}
    \caption{\label{figs3}  Density $n(\mu,T)$, normalized by the density $n_0(\mu,0)$ of the non-interacting Fermi gas at zero temperature and same chemical potential $\mu$, versus $(T/\mu)^2$. The solid blue line denotes the zero-temperature limit, $n(\mu,0)/n_0(\mu,0)=1/\xi^{3/2}$. The solid red line is a linear fit, resembling Fermi liquid behavior of the density above $T_c$.}
\end{figure}

\begin{figure}
    \centering
    \includegraphics[width=84mm]{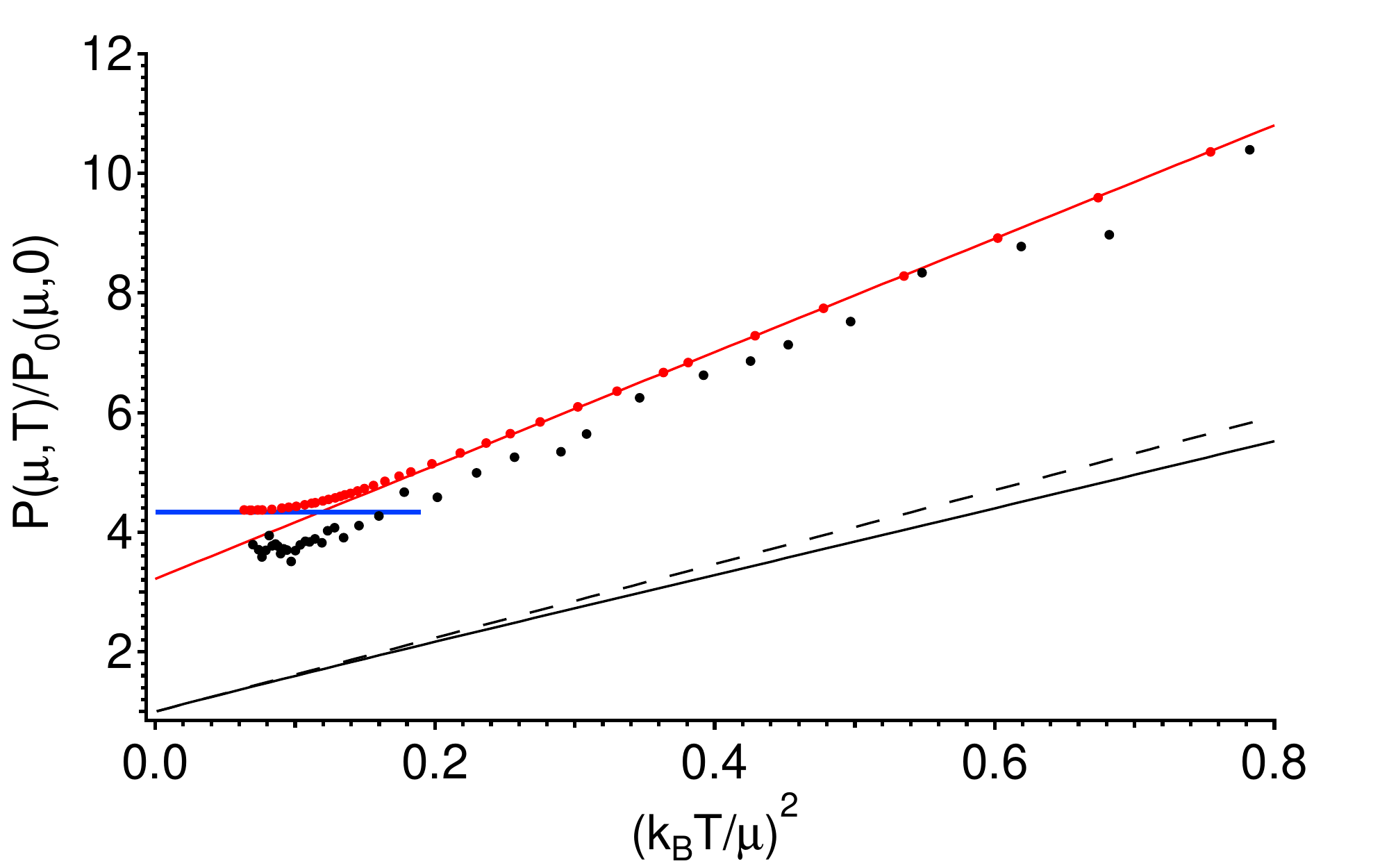}
    \caption{\label{figs4} Pressure $P(\mu,T)$, normalized by the pressure $P_0(\mu,0)$ of the non-interacting Fermi gas at zero temperature and same chemical potential $\mu$, versus $(T/\mu)^2$. Red solid circles are the experimental data from this work. Black solid circles are from~\cite{nasc09supp}. The solid blue line denotes the zero-temperature limit, $P(\mu,0)/P_0(\mu,0)=1/\xi^{3/2}$, the red solid line is a linear fit, resembling Fermi liquid behavior for the pressure above $T_c$. The black solid line shows the pressure of a non-interacting Fermi gas, the dashed solid line the linear approximation, valid if $T\ll T_F$.}
\end{figure}

At low temperatures, the normalized density reaches the zero-temperature value $1/\xi^{3/2}$.
The pressure does not show an upturn but smoothly attains a limiting value. The smooth behavior is expected for a second-order transition, where first derivatives of the pressure with respect to the chemical potential are continuous across the transition. The intercept of a straight line at the limiting value $P(\mu,0)/P_0(\mu,0) = 1/\xi^{3/2}$ with the linear fit at high temperatures underestimates $T_c/\mu_c$ by 22\%. This method resulted in $T_c/\mu_c = 0.32(2)$ in~\cite{nasc09supp} (shown for comparison), inconsistent with our determination. However, our value for $T_c/\mu_c = 0.40(3)$ agrees excellently with the most accurate theoretical determination $T_c/\mu_c = 0.400(14)$ from~\cite{goul10tcsupp}.
In a Fermi liquid, one has $P(\mu,T) = P_0(\mu,0)\left(\xi_n^{-3/2} + \frac{5\pi^2}{8}\xi_n^{-1/2}\frac{m^*}{m}\left(\frac{k_B T}{\mu}\right)^2\right)$~\cite{nasc09supp}. Although the agreement with the Landau Fermi liquid is fortuitous, we may nevertheless use it to model the data. From the linear fit versus $(T/\mu)^2$, we obtain the fit parameters $\xi_n = 0.46(1)$, in agreement with the determination of the same parameter from the chemical potential, and $\frac{m^*}{m} = 1.04(2)$.
Note that in the same temperature regime, a linear fit to the normalized pressure of a non-interacting Fermi gas (solid line in Fig. S4) gives $\frac{m^*_0}{m} = 0.91$, while clearly, this value for a non-interacting Fermi gas should not differ from unity. The reason for discrepancy is that the pressure of the non-interacting Fermi gas in the temperature range $0.2\apprle (T/\mu)^2 \apprle 1$ is not in the linear regime of low temperatures (shown as the dashed line in Fig. S4). Equivalently, the entropy per particle is not much smaller than $k_B$. The fit-parameter $m^*$ should thus not be taken literally as the effective mass of particles in the normal state. However, the general finding of~\cite{nasc09supp} of an only weakly renormalized normal state at unitarity remains true. Compared to the non-interacting Fermi gas in the same regime of $T/\mu$, the unitary gas has an apparent effective mass enhancement of only about 12\%.
In view of the specific heat, which is not linear in temperature above $T_c$, and the facts that neither $T/\mu$, nor the entropy per particle, are small above $T_c$, we conclude that the usual Landau Fermi liquid picture, valid when the temperature is much smaller than the Fermi energy, cannot properly describe all thermodynamic properties of the normal unitary Fermi gas. Note that our determination of fit parameters differs from the previous result~\cite{nasc09supp} $\frac{m*}{m} = 1.13(3)$ and $\xi_n = 0.51(2)$, possibly due to the mentioned differences in calibration between the experiments. This results in overall lower pressures (by a factor of about 13\%) than in the present experiment, where we find $\xi = 0.376(5)$, to be shifted by at most 2\% due to the error in the Feshbach resonance.





%

\begin{thebibliography}{10}
\bibitem{ku11supporting}%
  \BibitemOpen
  \bibinfo {note} {Materials and methods are available as supporting
  materials}%
  \bibAnnoteFile{NoStop}{ku11supporting}%
\end{thebibliography}
\end{document}